\documentstyle[11pt,twoside,paspconf,psfig,epsf]{article}

\def\deg   {$^\circ$}

\def\Pcos  {$\Phi^0$}

\def\tauLL {\tau_{\scriptscriptstyle LL}}

\def\nH    {n_{\scriptscriptstyle H}}
\def\Ha    {${\rm H}\alpha$}
\def\Lya    {${\rm Ly}\alpha$}
\def\eg    {{\it e.g.,\ }}
\def\ie    {{\it i.e.,\ }}
\def\qv    {{\it q.v.,\ }}
\def\etal  {{\it\ et al.}}
\def\kms   {\ km s$^{-1}$}
\def\intensity{\ifmmode{{\rm erg\ cm}^{-2}{\rm\ s}^{-1}
      {\rm\ Hz}^{-1}{\rm\ sr}^{-1}}
      \else {erg cm$^{-2}$ s$^{-1}$ Hz$^{-1}$ sr$^{-1}$}\fi}
\def\flux{\ifmmode{{\rm erg\ cm}^{-2}{\rm\ s}^{-1}}\else {erg
cm$^{-2}$ s$^{-1}$}\fi}
\def\fluxdensity{\ifmmode{{\rm erg\ cm^{-2}\ s^{-1}\ Hz^{-1}}}\else {erg
cm$^{-2}$ s$^{-1}$ Hz$^{-1}$}\fi}
\def\phoflux{\ifmmode{{\rm photons\ cm}^{-2}{\rm\ s}^{-1}}\else {photons
cm$^{-2}$ s$^{-1}$}\fi}
\def\vLMC {v_{\scriptscriptstyle\rm LMC}}

\def\gtrapprox{\;\lower 0.5ex\hbox{$\buildrel >\over \sim\ $}}
\def\lessapprox{\;\lower 0.5ex\hbox{$\buildrel < \over \sim\ $}}

\def\deg   {$^\circ$}

\def\Pcos  {$\Phi^0$}

\def\tauLL {\bar\tau}

\def\nH    {n_{\scriptscriptstyle H}}

\def\Ha    {${\rm H}\alpha$}
\def\eg    {{\it e.g.,\ }}
\def\ie    {{\it i.e.,\ }}
\def\cf    {{\it cf.\ }}
\def\qv    {{\it q.v.,\ }}
\def\etal  {{\it\ et al.}}
\def\intensity{\ifmmode{{\rm erg\ cm}^{-2}{\rm\ s}^{-1}
      {\rm\ Hz}^{-1}{\rm\ sr}^{-1}}
      \else {erg cm$^{-2}$ s$^{-1}$ Hz$^{-1}$ sr$^{-1}$}\fi}

\def\Em{\ifmmode{{\rm E}_m}\else {{\rm E}$_m$}\fi}
\def\Nh{\ifmmode{{\rm N}_{\scriptscriptstyle\rm H}}\else {{\rm N}$_{\scriptscriptstyle\rm H}$}\fi}

\def\phiIV{\ifmmode{\varphi_4}\else {$\varphi_4$}\fi}
\def\phiI{\ifmmode{\varphi_i}\else {$\varphi_i$}\fi}
\def\Dm{\ifmmode{{\cal D}_m}\else {{\cal D}$_m$}\fi}
\def\fesc{\ifmmode{f_{\rm esc}}\else {$f_{\rm esc}$}\fi}
\def\rsolar{\ifmmode{r_\odot}\else {$r_\odot$}\fi}
\def\emunit{\ifmmode{{\rm cm}^{-6}{\rm\ pc}}\else {
cm$^{-6}$ pc}\fi}
\def\Dm{\ifmmode{{\cal D}_m}\else {{\cal D}$_m$}\fi}
\def\fesc{\ifmmode{f_{\rm esc}}\else {$f_{\rm esc}$}\fi}
\def\rsolar{\ifmmode{r_\odot}\else {$r_\odot$}\fi}
\def\emunit{\ifmmode{{\rm cm}^{-6}{\rm\ pc}}\else {
cm$^{-6}$ pc}\fi}
\def\flux{\ifmmode{{\rm erg\ cm}^{-2}{\rm\ s}^{-1}}\else {erg
cm$^{-2}$ s$^{-1}$}\fi}
\def\fluxdensity{\ifmmode{{\rm erg\ cm^{-2}\ s^{-1}\ Hz^{-1}}}\else {erg
cm$^{-2}$ s$^{-1}$ Hz$^{-1}$}\fi}
\def\phoflux{\ifmmode{{\rm phot\ cm}^{-2}{\rm\ s}^{-1}}\else {phot
cm$^{-2}$ s$^{-1}$}\fi}
\def\phorate{\ifmmode{{\rm\ phot\ s}^{-1}}\else {\ phot s$^{-1}$}\fi}

\begin{document}

\title{The Galactic Halo Ionizing Field and H$\alpha$ Distances to HVCs}

\author{J. Bland-Hawthorn}
\affil{Anglo-Australian Observatory, PO Box 296, Epping, NSW 2121, Australia}
\author{P.R. Maloney}
\affil{Center for Astrophysics \& Space Astronomy, University of Colorado, Boulder, CO 80309-0389}

\begin{abstract}
  There has been much debate in recent decades as to what fraction of
  ionizing photons from star forming regions in the Galactic disk
  escape into the halo. The recent detection of the Magellanic Stream
  in optical line emission at the CTIO 4m and the AAT 3.9m telescopes
  may now provide the strongest evidence that at least some of the
  radiation escapes the disk completely. While the distance to the 
  Magellanic Stream is uncertain, the observed \Ha\ emission is most 
  plausibly explained by photoionization due to hot, young stars. 
  Our model requires that the mean Lyman-limit opacity perpendicular to 
  the disk is $\tauLL \approx 3$, assuming the covering fraction of the 
  resolved clouds is close to unity. Within the context of this model, 
  it now becomes possible to determine distances to high velocity clouds, 
  and the 3D orientation of the Magellanic Stream. Here, we discuss
  complications of the model (\eg porosity, topology), future tests,
  ongoing improvements, and the importance of \Ha\ limb brightening from 
  surface ionization. More speculatively, we propose a direct experiment
  for locating an HVC in 6-dimensional phase space above the Galactic 
  plane.
\end{abstract}

{\bf Keywords:}
interstellar medium $-$ intergalactic medium $-$ individual object: 
Magellanic Stream $-$ Galaxy: corona, halo $-$ interferometry

\section{Introduction}
After two full days at this first ever {\sl High Velocity Clouds} (HVCs)
workshop, there were portents that we are on the threshold of a 
Renaissance in the study of the interstellar and the 
intergalactic medium. Space-borne UV spectroscopy reveals nearby counterparts
to distant \Lya\ clouds (Sembach; see Sembach\etal\ 1998). HVCs with high HI 
columns are now routinely detected in \Ha\ emission (Tufte, Reynolds \& 
Haffner 1998; Bland-Hawthorn\etal\ 1998). The exquisite 
{\tt HIPASS} observations of the Stream reveal great complexity (Putman;
see Putman\etal\ 1998).  Blitz proposed a far-reaching scenario that 
many HVCs are hundreds of kiloparsecs away, and are evidence for debris 
from the formation of the Local Group. 

In collaboration with Putman \& Gibson, we find that most, if not all, 
resolved gas clouds observed to date $-$ with a mean column density 
higher than 10$^{19}$ cm$^{-2}$ $-$ can be detected in sensitive \Ha\ 
observations.  Of course, this statement comes with caveats emptor since
the expected threshold for which this holds true is an order of magnitude
lower. However, there was widespread recognition of the importance of direct 
\Ha\ detections of HVCs (for a review, see Wakker \& van Woerden 1998; 
Tufte\etal\ 1998; Bland-Hawthorn\etal\ 1998). We shall argue that \Ha\ 
detections towards high column HI structures can be converted to distances 
if these clouds are situated in the halo of the Galactic ionizing field.
If correct, \Ha\ detections 
{\it and non-detections} are set to revolutionize the field.

Here, we discuss potential pitfalls of the distance model, future
improvements, and complicating factors like cloud topology and covering
fraction.

\section{The escape of ionizing photons from the Galaxy}
There has been extensive theoretical and observational
interest in establishing what fraction of the total ionizing
luminosity from the stellar disk of the Milky Way and other galaxies
escapes into the halo and the intergalactic medium (\eg Miller \& Cox
1993; Dove \& Shull 1994; Leitherer \& Heckman 1995). Diffuse ionized
gas between HII regions in half a dozen well studied galaxies suggests
that a significant fraction escapes to ionize the ambient ISM (e.g.
Hoopes, Walterbos \& Greenawalt 1996; Ferguson\etal\ 1996; 
Greenawalt\etal\ 1998). Broadly
speaking, if the optical depth at the Lyman limit is $\tauLL$, these
observations require $\tauLL\approx 1$ on scales of $\sim$100 pc.
The vertically extended Reynolds layer requires that $\tauLL \approx 2$ 
to explain the observed line emission (Reynolds 1990). At the distance
of the Magellanic Stream, we have shown (Bland-Hawthorn \& Maloney 1999)
that the observed \Ha\ emission measures (Weiner \& Williams 1996, hereafter
W$^2$) are consistent with ionization 
by the Galactic disk providing 6\% of the ionizing radiation ($\tauLL 
\approx 3$ perpendicular to the disk) escapes into the Galactic halo. 
The influence of this residual radiation is easily detected with sensitive 
\Ha\ observations out to great distances (100 kpc or more) from the Galaxy.

\section{The shape of the halo ionizing field}
The emission measure \Em\ from the surface of a cloud
embedded in a bath of ionizing radiation gives a direct gauge,
independent of distance, of the ambient radiation field beyond the
Lyman continuum (Lyc) edge (\eg Hogan \& Weymann 1979).  If the strength 
and direction of the radiation field is known, the observed \Ha\ emission 
from the surface of the cloud can be used to determine the cloud distance.
This assumes that the covering fraction ($\kappa$) and the projected
cloud topology $-$ {\it seen by the ionizing photons} $-$ is known and that 
there are sufficient gas atoms to soak up the incident ionizing photons.  
In a spate of recent papers (Bland-Hawthorn \& Maloney 1997, 1999; 
Bland-Hawthorn\etal\ 1998), we develop an idealized, opaque disk $+$
halo model for predicting the \Ha\ emission measure at an arbitrary
point within the Galactic halo. The reader may investigate this model 
through the {\tt pgperl} web tool at 
\begin{quote}
{\tt http://www.aao.gov.au/cgi-bin/diskhalo} .
\end{quote}
Fig.~1 shows how the surfaces of constant ionizing flux (in units
of \phiIV, \ie 10$^4$ ionizing photons cm$^{-2}$ s$^{-1}$) are
highly elongated, producing the appearance of `ionization cones', in the 
direction of the spin axis above an opaque disk. This is consistent with 
the constant ionization seen along
extended filaments perpendicular to the Galactic disk (Haffner, Reynolds
\& Tufte 1998) and starburst disks (\eg Phillips 1993; Shopbell 1996).
The dotted lines illustrate the possible influence of the LMC.
Without the LMC, the ionizing field is axisymmetric. For clouds that
lie within 10 kpc or so of the disk plane, the model needs to include
the known positions of the spiral arms (see below).

\psfig{file=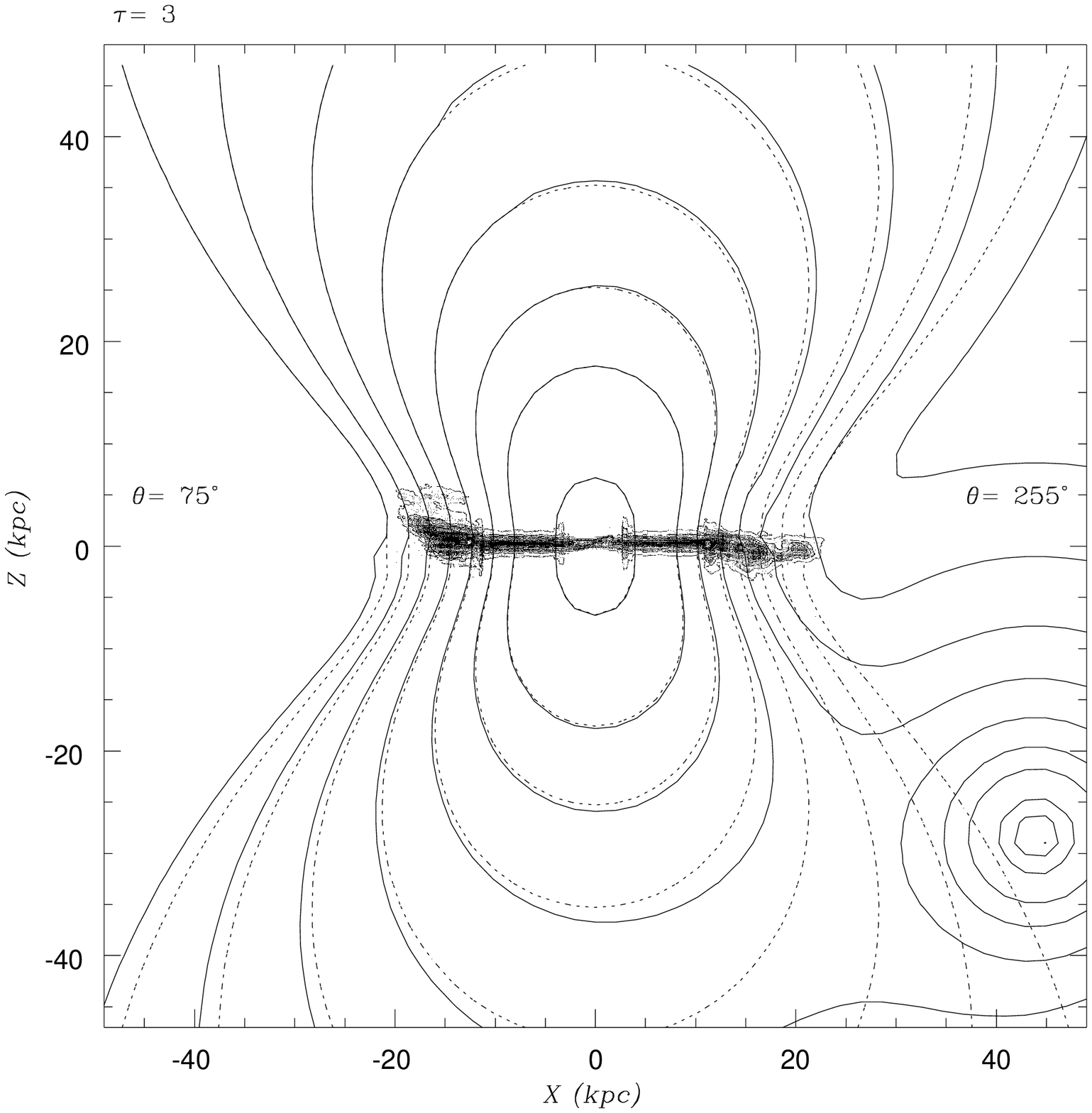,width=13cm}

\smallskip\noindent {\sl Fig.~1.} The Galactic halo ionizing field. 
The coordinates are
with respect to a plane perpendicular to the Galactic disk (with
the Galactic centre at the origin) at a constant galactic azimuth
angle (75\deg, 255\deg). The dotted lines show the ionizing flux
($\varphi_4$) due to the galactic disk; the solid lines include the
contribution from the LMC ($1.5\times 10^{52}\phorate$). 
The opacity of the HI disk (shown in
half tone) has been included. The contours, from outside in, are
for $\log \varphi_4 = 1, 1.25, 1.5, 1.75, 2, 2.25, 2.5, 3$ phot cm$^{-2}$
s$^{-1}$. The minor contribution from the Galactic corona is omitted.

\section{Tests of the Galactic ionizing field model}

Our working model for the Magellanic Stream \Ha\ detections requires
independent confirmation.  The disk-halo ionization model implicitly 
assumes that O (and B) stars dominate the halo energetics out to great 
distances.  For the model to survive, it must pass a number of important 
tests.

\subsection{Dilute \& shock ionizing radiation fields}

The emission line diagnostics for stellar photoionization are fairly
well understood (Fig.~2).
The ionizing spectrum of an O star drops precipitously beyond 20 eV 
(Kurucz 1979; Vacca\etal\ 1996), with 10\% of the energy emerging in the 
interval 24.6 eV to 54.4 eV, and only a tiny fraction appearing beyond the 
He II edge at 54 eV (however, \cf Schaerer \& de Koter 1997). This tapering
energy band is only able to excite a limited range of optical diagnostics 
in the strong field limit, with an even more restricted list in the weak
field limit.

Large-scale shocks as a source of ionization are an attractive
prospect, particularly in light of the spectacular bow-shock
morphology at the head of, and in small knots along, the Magellanic
Stream (Mathewson \& Ford 1984). However, such a morphology
is much less evident in the {\tt HIPASS} survey (Putman\etal\ 1998).
The shocks one anticipates in conventional drag models are unlikely to
produce significant amounts of \Ha\ emission. The radiative regions in
shocks are in pressure equilibrium with the external gas (Sutherland
\& Dopita 1993) such that
\begin{equation}
n_{\scriptscriptstyle\rm A} \vLMC^2
\approx n_{\scriptscriptstyle\rm S} v_{\scriptscriptstyle\rm S}^2
\end{equation}
where $n_A$ is the ambient density, $\vLMC$
is the speed of the Stream in the frame of the Galaxy, 
$v_{\scriptscriptstyle\rm S}$ and $n_{\scriptscriptstyle\rm S}$ 
are the shock velocity and the post-shock density. We adopt a coronal
density of $n_{\scriptscriptstyle\rm A} \approx 10^{-4}$, 
the maximum allowed by pulsar 
dispersion measures; at the head of clouds MS II$-$IV, the 
volume-averaged atomic density from the {\tt HIPASS} observations 
is in the range $n_{\scriptscriptstyle\rm S} = 0.1-1$ cm$^{-3}$.
The \Em\ values quoted by Weiner \& Williams produce electron densities 
in our range for any reasonable path length. Proper motion studies indicate 
that the total Galactocentric transverse velocity for the LMC is 
$\vLMC = 213\pm 49$\kms\ (Lin, Jones \& Klemola 1995).
The predicted shock velocities arising from the Stream dynamics are 
only a few \kms, which are far too small to ionize hydrogen.

There exists another class of shock models, not discussed by 
Weiner \& Williams, which could conceivably ionize the clouds. Gas 
heated by supernovae can puncture the cold
disk and escape into the halo (\eg Norman \& Ikeuchi 1989).  The hot halo
gas may form a galactic wind, a hydrostatic atmosphere around the
Galaxy or, more likely, it may eventually cool, recombine and descend
onto the disk, establishing a galactic fountain (Shapiro \& Field
1976; Bregman 1980; Houck \& Bregman 1990; Shapiro \& Benjamin 1994).
In this picture, the Galactic fountain is largely responsible for the
hot corona and may possibly explain HVCs as material involved in the
circulation process (but see Wakker \& van Woerden 1991).  The cool, 
descending gas
of the galactic fountain would be detectable at anomalous velocities
in 21 cm emission.  Theoretically, the maximum velocity of gas
streaming to the disk is roughly 100 km s$^{-1}$.  High apparent
velocities are predicted only for distant clouds, because of the
particular mechanism of acceleration in combination with projection
effects induced by galactic rotation.  Thus, the model can explain the
existence of gas with velocities up to 200 km s$^{-1}$, but not if it
is generally nearby.

\medskip
\psfig{file=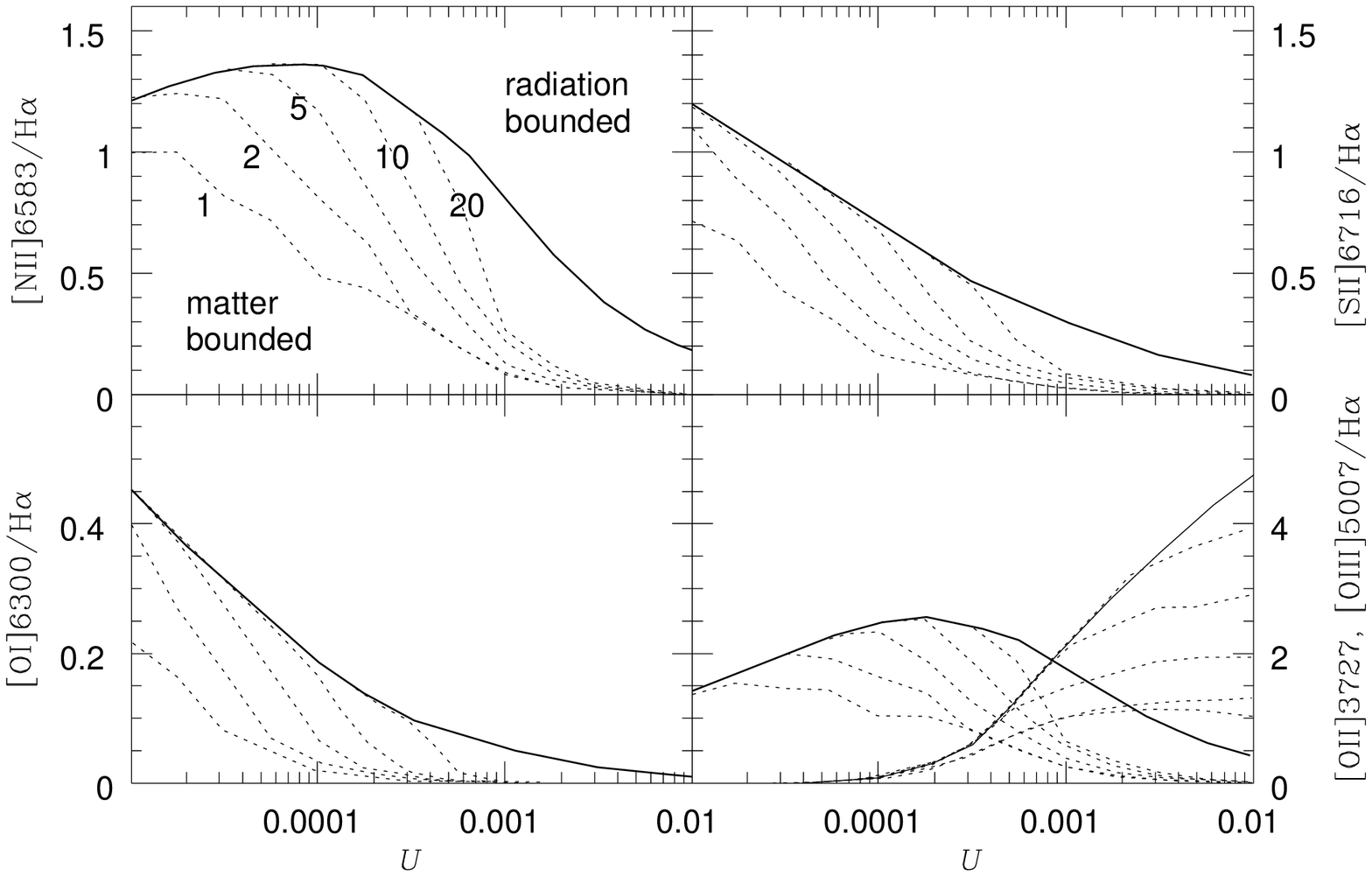,width=13cm}

\smallskip\noindent {\sl Fig.~2.} 
The dependence of five important line diagnostics on the ionization
parameter $U=\varphi_i/c \nH$.  The solid lines 
are radiation bounded models where all ionizing photons are soaked up by
the gas. Also shown are matter bounded models where the slab has been
truncated at 1, 2, 5, 10 and 20 $\times$ 10$^{18}$ cm$^{-2}$. 

\subsection{Spectral diagnostics}

The importance of detecting HVCs in more than one emission line
cannot be overstated.  Simultaneous detections have already been 
achieved with both Fabry-Perot and faint object spectrographs 
(Bland-Hawthorn\etal\ 1998; 1999). For example,
a fountain is expected to produce strong low ionization lines 
([OI], [OII], [NII], etc.) due to shocks and perhaps
metal-enriched material from the central disk (Shapiro \& Benjamin
1993). Unfortunately, these are the same diagnostic lines produced
by dilute stellar photoionization, but multi-line spectroscopy 
remains a crucial probe of the radiation field. The discovery of, say,
strong [OIII] emission anywhere along the Magellanic Stream would
indicate that some unforeseen ionization/heating mechanism is at work,
and would cast doubt on our disk ionization model. 

For a gas with known metal abundances, line diagnostics can be used
to constrain
the ionization parameter ${\cal U} = \varphi_i / c \nH$.  The few HVCs 
with constrained metallicities appear to have abundances no more than 
one third of solar (de Boer \& Savage 1984; Blades\etal\ 1988).  
The most commonly used line ratios, when plotted as a function of 
$\cal U$, are characterized by bell-shaped curves with different first and 
second moments (see Fig.~2). The decline at high and low $\cal U$ marks the 
change in electron temperature and the increasing importance of higher and 
lower ionization states, respectively. Detailed line
ratios have been computed for a range of ionizing sources, including
shocks, non-thermal power laws, and hot young stars observed through
an opaque medium (Mathis 1986; Veilleux \& Osterbrock 1987; Sokolowski
1992, 1995; Sutherland \& Dopita 1993; Shapiro \& Benjamin 1993). 
As Fig.~2 shows, a complicating factor is whether the gas clouds are
matter or radiation bounded.

The limited broadband spectroscopic work to date on HVCs implicates
dilute photoionization. We did not detect [OIII] towards MS~II~A or
the Smith Cloud, and in the latter study, [NII]/\Ha\ was observed to be 
of order unity (Bland-Hawthorn\etal\ 1998; 1999). Furthermore,
a recent spectrum with the MSO 2.3m double beam spectrograph shows 
evidence for a similar enhancement of [SII]/\Ha\ towards MS~II~A. 
Taken together with the close match between the \Ha\ and HI kinematics,
the spectral diagnostics (Fig.~2) are suggestive of static photoionization 
of cloud surfaces in a very dilute, ionizing field.

\section{\Ha\ distances to HVCs}

We anticipate that 
HVCs will be detectable in sensitive \Ha\ observations to distances of
more than 100 kpc. Several HVCs have now been detected in \Ha\
emission (Kutyrev \& Reynolds 1989; M\"unch \& Pitz 1990;
Bland-Hawthorn\etal\ 1998; Tufte, Reynolds, \& Haffner 1998), at levels of
$\approx0.1-0.3$ R, in addition to the W$^2$ Magellanic Stream
detections. A detailed discussion of the
use of \Ha\ observations as a distance indicator for HVCs is presented
in Bland-Hawthorn\etal\ (1998). Here we simply note that: (1) although there
are likely to be difficulties with inferring the distances for clouds
within $z\lessapprox 10$ kpc above the plane, due to structure in both
the absorption and ionizing photon emissivity on smaller scales, our
ionizing field model predicts the
distances of the M and A complexes (as observed by Tufte\etal\ 1998)
to within a factor of two; (2) by mapping the emission measure along
the Magellanic Stream, it will be possible to
infer both the actual three-dimensional geometry of the Stream with
respect to the Galactic plane, and the value of \fesc\ for the LMC,
which may produce a modest but measurable increase in the emission
measure in the Stream near the LMC.

\section{Distance uncertainties}

\subsection{Primary calibrators}

In practice, we do not doubt that the uncertainties in \Ha\ distances 
to HVCs are large
(\ie factors of a few), particularly for clouds close to the plane.
Our models assume that the covering fraction of the HI slab
is unity as seen by the ionising photons, and that the Galactic
dust and ionising sources are smoothly distributed. But we would argue
that crude distances, and even the ability to distinguish between
`nearby' (detection) and `far away' (non-detection), is a major advance.
Even without a reliable primary distance calibrator, there is a great 
deal to learn from relative distance measurements, although this assumes
that clouds possess the same basic structure and topology regardless of
size. The Magellanic Stream may constitute the best prospect for a 
primary calibrator, if the distance to a particular cloud along the Stream
could be determined independently and unambiguously. In the foreseeable 
future, the prospects for an independent calibration look grim.

\subsection{Porosity}

A problem that has haunted us for several years now refuses to go away:
\begin{quotation}
{\sl For a sheet of gas lying in the plane of the sky, can an observer 
determine experimentally what fraction of the projected area is opaque
to Lyman limit photons?}
\end{quotation}
This problem is intimately linked to a concern raised by Bregman during the
meeting. The HI maps of Mathewson \& Ford (1984) give the impression that 
atomic gas forms a continuous screen along the Stream with a column density
exceeding the Lyman limit threshold. If this was true, then the weakest
\Ha\ measurements, or even the non-detections, are more relevant for 
normalizing the Galactic ionizing field; the stronger detections would
require an independent source.

The HI covering fraction ($\kappa$)\footnote{$\kappa$ is sometimes
referred to as the {\sl covering factor} or the {\sl beam filling fraction}.
The covering fraction and volume filling fraction
$f$ are linearly related in the limit of small $\kappa$, although this
no longer holds true as $\kappa$ and $f$ approach unity.}
remains the major stumbling block to using clouds as sensitive probes of 
the ambient radiation field. Simply put,
this is the fraction of area within unit solid angle that is taken up
by gas with a hydrogen column density sufficient to absorb ionizing
photons below a specific energy. Since the absorption cross-section
declines rapidly with energy, we shall restrict our definition to
photons with energy less than or equal to 250 eV.

The covering fraction seen from Earth is much
less important than the covering fraction seen by the ionizing
photons.  A case in point is the HI 1225+01 cloud discovered by
Giovanelli \& Haynes (1991) in the Local Supercluster. This has been
used by various authors (\eg Vogel\etal\ 1995) to set limits
on \Pcos, the metagalactic UV flux. However, the structure is now 
thought to be flattened and highly inclined to the line of sight
(Chengalur, Giovanelli \& Haynes 1995). 

To be sure of even the global structure, an HI screen should have its 
longest dimensions close to the plane of the sky. Even under ideal 
circumstances, it is difficult to determine $\kappa$ unambiguously. 
A smooth HI
distribution gives essentially no structural information on scales
smaller than the beam size. A large variance in the structure
(\eg Dickey 1979) compared to the mean HI column might imply a lumpy
distribution although temperature fluctuations and/or kinematic
variations can produce the same effect. HI clouds may comprise dense
cores and an ambient diffuse component (\eg Ferrara \& Field
1993). It may be possible to use extended baseline, 21 cm spectroscopy
to resolve the clumps on small ($\sim$0\farcs 1) scales. A comparison
with single dish observations will establish whether the remaining
gas is opaque to ionizing photons if smoothed over the remaining solid
angle between the dense clumps.

There has been extensive work on Galactic HI clouds by looking for
absorption towards background radio continuum sources on small angular
scales (Payne, Salpeter \& Terzian 1983; Colgan, Salpeter \& Terzian
1990). Direct observations of the HI emission are complicated by the
need for both radio interferometers and single dish observations
(Crovisier \& Dickey 1983). In single dish observations, it is
straightforward to produce an autocorrelation function (ACF) of the
observations. But whether this constitutes a useful probe of
statistical variations is doubtful, particularly with the regular
occurrence of side-lobe contamination. Unfortunately, most authors
(\eg Crovisier \& Dickey 1983) publish the ACF after subtracting the
mean and normalizing to unit variance without stating these
quantities. Alternatively, interferometers measure the complex
visibility function along particular tracks in $u-v$ space, which can
be converted to a fringe amplitude and therefore a power spectrum.
The zero spacing is never observed which complicates the process of
normalizing the higher resolution interferometric data with single
dish observations.

For Galactic HI clouds, ROSAT HI shadows may provide the best clues,
and appear to indicate a covering fraction close to unity (Snowden
\etal\ 1991; Burrows \& Mendenhall 1991). Here, a comparison is
made between IRAS or HI observations and ROSAT observations at each
independent beam position.  Since soft X-ray emission is absorbed by
neutral atomic gas, the known $\Nh$ (for an assumed dust to gas ratio)
can be compared with the fraction of X-rays that gets through compared
to the background.  The method has its weaknesses: (i) the slope of
the X-ray spectrum is assumed, (ii) the HI observations are smoothed
to the relatively poor beam size of the ROSAT observations.

More speculatively, by analogy with ROSAT shadows, it may be possible
(cf. Gould \& Sciama 1964) to determine $\kappa$ for gas clouds projected
against hot cluster gas or even attenuation of unresolved X-ray sources in
deep ROSAT observations (Fabian \& Barcons 1992).
At optical wavelengths, there are only a few quasars per square
degree bright enough for line spectroscopy. Deep Keck images at B$=$
26 mag arcsec$^{-2}$ show that up to one million galaxies are to be found 
within each square degree (Smail\etal\ 1995). A comparison of the
number counts observed through an HI cloud compared to an offset field 
could be used to establish the mean opacity through the disk. However, 
the faintest sources may be highly clustered on large scales which would 
complicate the analysis.

Detailed mapping of HI clouds in several emission lines (\eg\ Fig.~2) may 
go some way to answering the question posed at the beginning of this 
section.  It is difficult to identify a rigorous, definitive test: the 
eventual resolution may come down to a circumstantial body of evidence. 
There is, however, another motivation for optical line maps, as we now 
discuss.

\section{The importance of limb brightening}

Any form of limb brightening observed in optical/IR emission lines
near HVCs, and HI structures in general, has important ramifications.
In Fig.~3, the expected contrast level ${\cal C}$ is given roughly by
\begin{equation}
\label{limb}
{\cal C} \sim 2 \left({d{\cal R}}\over{\cal B}\right)^{0.5}
\left({\cal R}\over{\cal B}\right)^{0.5}
\end{equation}
where $d{\cal R}$ is the depth of the ionized layer in parsecs, ${\cal R}$ is 
the radius of curvature at the ionized surface, and ${\cal B}$ is the beam size
of the spectrograph. The skin depth can be estimated crudely from (i) the
observed emission measure ($d{\cal R} \sim \Em n^{-2}_e$) along a sight line
through the middle of the cloud, 
(ii) the inferred radiation field ($d{\cal R} \sim 10^{-1.9} \phiIV n^{-2}_e$), 
or (iii) the inferred ionization parameter
($d{\cal R} \sim 10^{6.3} {\cal U} n^{-1}_e$), 
assuming the mean density has not changed after ionization.
Clearly, the degree of contrast will depend on the 
beam size, \ie\ 1\deg, WHAM; 0.5\deg, Las Campanas; $<$0.15\deg\ 
for most other groups. As many clouds as possible should be observed by
both small and large beams.

\medskip\bigskip
\begin{minipage}{6cm}
\psfig{file=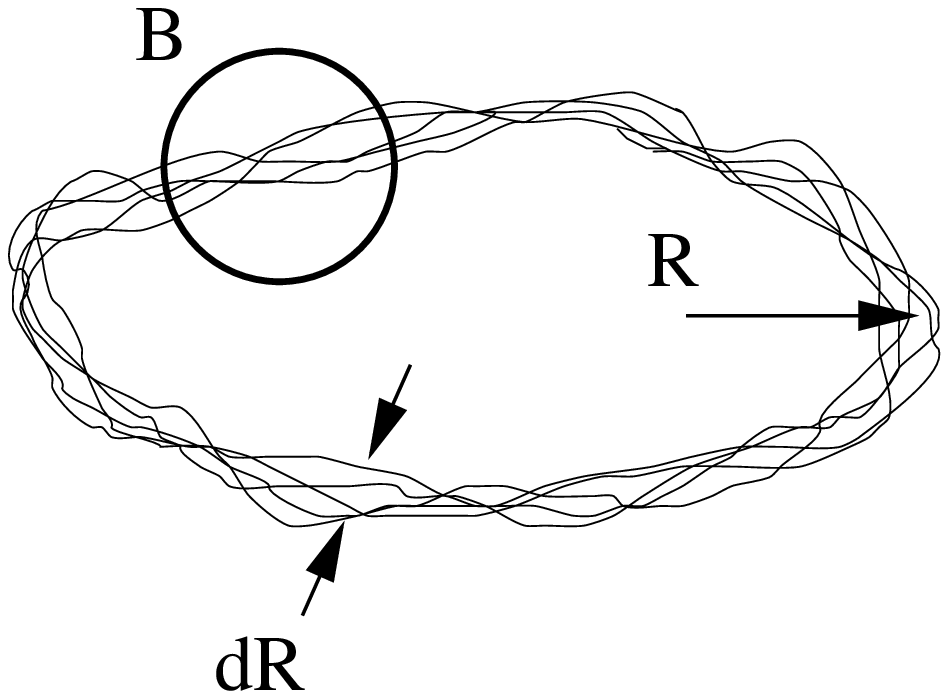,width=6cm}
\end{minipage}
\qquad
\begin{minipage}{6cm}
\noindent {\sl Fig.~3.} The instrument beam size ${\cal B}$
projected onto the limb brightened edge of a high velocity cloud.
\end{minipage}

\medskip\bigskip
The presence of limb brightening can tell us many things about the cloud
and the ionizing source:
\begin{itemize}
\item The cloud has a sharp edge and is dense enough to stop ionizing
photons within a short stopping length.
\item The volume filling fraction $f$ of the cloud is finite, say, 
$f > 10^{-2}$, as we demonstrate with the fractal model below.
\item The brightest \Ha\ emission may trace the overall morphology of the 
illuminated cloud.
\item If the limb brightening is restricted to one side of the cloud, this
gives the direction of the dominant ionizing source, or the direction of 
motion in the case of shocks; the emission measure from the opposite side
provides a sensitive gauge of the metagalactic UV flux.
\item It is possible that the degree of contrast alone is a crude distance
indicator, \eg WHAM is only expected to see high contrasts for
nearby clouds.
\end{itemize}

\section{Topology: complex clouds and cloud complexes}

A major concern is that clouds do not constitute simple structures on
the scale size of the beam. This is a well known problem in atmospheric
physics, particularly in studies of terrestrial clouds.  Wakker raised 
the matter of how HI structures should be classified. Statistical
estimators (\eg N-point correlators) are sensitive to gaussian fluctuations, 
multifractals, Markov chains, hierarchies, random walks, and so on. 
The choice of estimator largely reflects the physical process under
study, which is largely unknown for structure in HI surveys. Wavelet 
analysis and topological measures (\eg Minkowski functionals) have been 
shown to work well with noisy pixelated data (\eg Hobson, Jones \& Lasenby 
1998). 

\medskip
\psfig{file=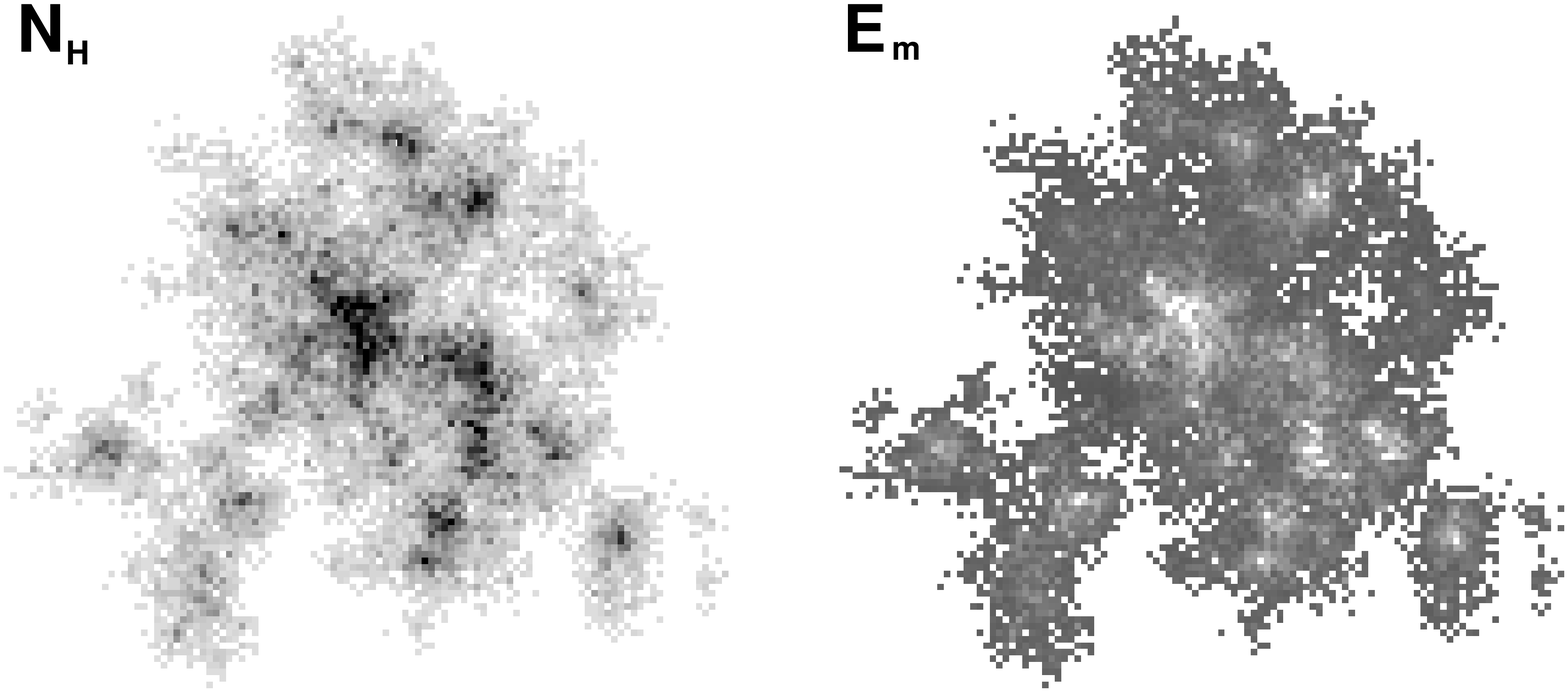,width=13cm}

\noindent {\sl Fig.~4.} A fractal cloud of dimension ${\cal D} = 3$
comprising 8000 cells, each with column density $3\times 10^{17}$ cm$^{-2}$,
ionized by an isotropic background.  The projected column density of the 
cloud (left) shows many volume brightened regions,
compared to the projected emission measure (right) which shows limb 
brightening on the smallest and largest scales.

\medskip\bigskip
A common starting point is to adopt a `building block' associated
with the simplest detectable structure at high resolution. More
specifically, in eight pointings with our 10\arcmin\ TAURUS beam towards 
clouds identified by {\tt HIPASS}, \Ha\ emission is always detected along
sight lines where the beam-averaged $\Nh > 10^{19}$ cm$^{-2}$. (For the
time being, we choose to overlook the fact that this threshold is an order 
of magnitude higher than expected for UV photons.) Let us suppose that
columns higher than our threshold arise from compact cloud structures
within the Stream.

We have examined cloud geometry and porosity for a range of distributions 
(\eg fractals; see Pfenniger \& Combes 1994).
The fractal cloud in Fig.~4 has an inverse square density law, and the
surface of the cloud is irradiated by an isotropic ionizing background.
Our simulations produce clouds that are roughly spheroidal, although this
is not a restriction (Voss 1985).  We find that `volume 
brightened' clouds\footnote{We adopt `volume brightened' as the 
antonym of `limb brightened' to indicate that spheroidal clouds appear 
brighter through the centre due to the larger column.} at 21 cm, with large 
filling factors, can exhibit `limb
brightening' at \Ha\ on arcminute scales.
Note that the HI cloud appears to have a dense core and an outer 
envelope (cf. Ferrara \& Field 1994).

In Fig.~4, the left and right maps appear to be inversely related, 
but this is only a property of some specific fractal models, and is not 
generally true.  Imagine a sphere which we begin to populate
uniformly but sparsely with optically thick cloudlets. The projected
\Ha\ emission measure and HI column density are both volume brightened.
As the number of cloudlets increases, we pass through an interesting
transition where the \Ha\ emission is now uniform over the cloud while
the HI remains volume brightened. This arises because the outer cloudlets
partially shadow the inner cloudlets from the external field. In the limit 
of large $f$, the \Ha\ emission becomes limb brightened, in contrast to the 
HI which remains volume brightened.

For a large set of matched HI and \Ha\ pointings,
much can be learnt from simple 2D histograms with axes
(\Nh,\ \Em). For example, any sort of relation where \Em\ is proportional 
to some positive power of $\Nh > 3\times 10^{17}$ cm$^{-2}$ would indicate 
$\kappa \ll 1$.  Our fractal clouds produce histograms with exponential 
slope inversely related to \Nh, where the 
slope is related to ${\cal D}$, which can be understood analytically.
A complete theoretical analysis must consider clouds
both near and far from the ionizing source {\it and} the observer. 
The surface ionization of clouds close to the disk will be highly
susceptible to the poorly known distribution of dust and UV sources,
and maybe shadowing of the disk by other HVCs. 

There may be useful information within individual \Ha\ exposures.
For clouds detected at the level of 1R, it should be possible to obtain 
useful detections after dividing a single exposure into discrete segments.
Since the radiation field at large galactocentric distances changes smoothly, 
large \Ha\ fluctuations across individual clouds and along the Stream are 
indicative of clumpiness.  If all clouds were identical in structure, the 
mean \Ha\ signal and \Ha\ variances constitute {\it independent} distance 
estimators.  

In summary, the \Ha\ variances, the degree of limb brightening and the
mean flux level, taken together, are needed to improve the reliability
of the \Ha\ distance method.

\section{Improvements to the Galactic halo ionization model}

The Galactic halo ionization model is discussed in detail elsewhere
(\qv Bland-Hawthorn\etal\ 1998). In the long term, its general 
acceptance will be intimately linked to the interpretation of the 
Magellanic Stream \Ha\ detections.  In the short term, there are
two burning issues to be addressed: (i) the escape of UV photons 
from the LMC (which bears on our derived \fesc\ for the Galaxy), 
and (ii) the incorporation of spiral arms into the disk model to 
improve distances to clouds within 10 kpc of the Galactic plane.

\subsection{The escape of UV photons from the LMC}

The Magellanic Stream and the Magellanic Clouds appear to share a
common link. The current understanding is that the Clouds are in a
close binary orbit about the Galaxy (Murai \& Fujimoto 1980) and are
engulfed in a common HI envelope (Kerr\etal\ 1954; Westerlund 1990).
The Magellanic Stream appears to extend from the Lagrangian point
between the Clouds and is observed to circle the Galaxy in an inclined
plane with respect to the disk (although see Lin, Jones \& Klemola
1995). The LMC has several highly active star forming regions,
particularly regions of very recent star formation (Shapley III) and
of ongoing star formation (30 Doradus).  Fujimoto \& Sofue (1976) give
a specific position for the LMC in Galactic coordinates, $(X,Y,Z) =
(-43,2,-28)$ kpc, which is close to the plane $Y=0$.  In Fig.~5, we
show the influence of the LMC on the poloidal radiation field in the
$X$-$Z$ plane passing through the Galactic Center.

\medskip\bigskip
\centerline{
\psfig{file=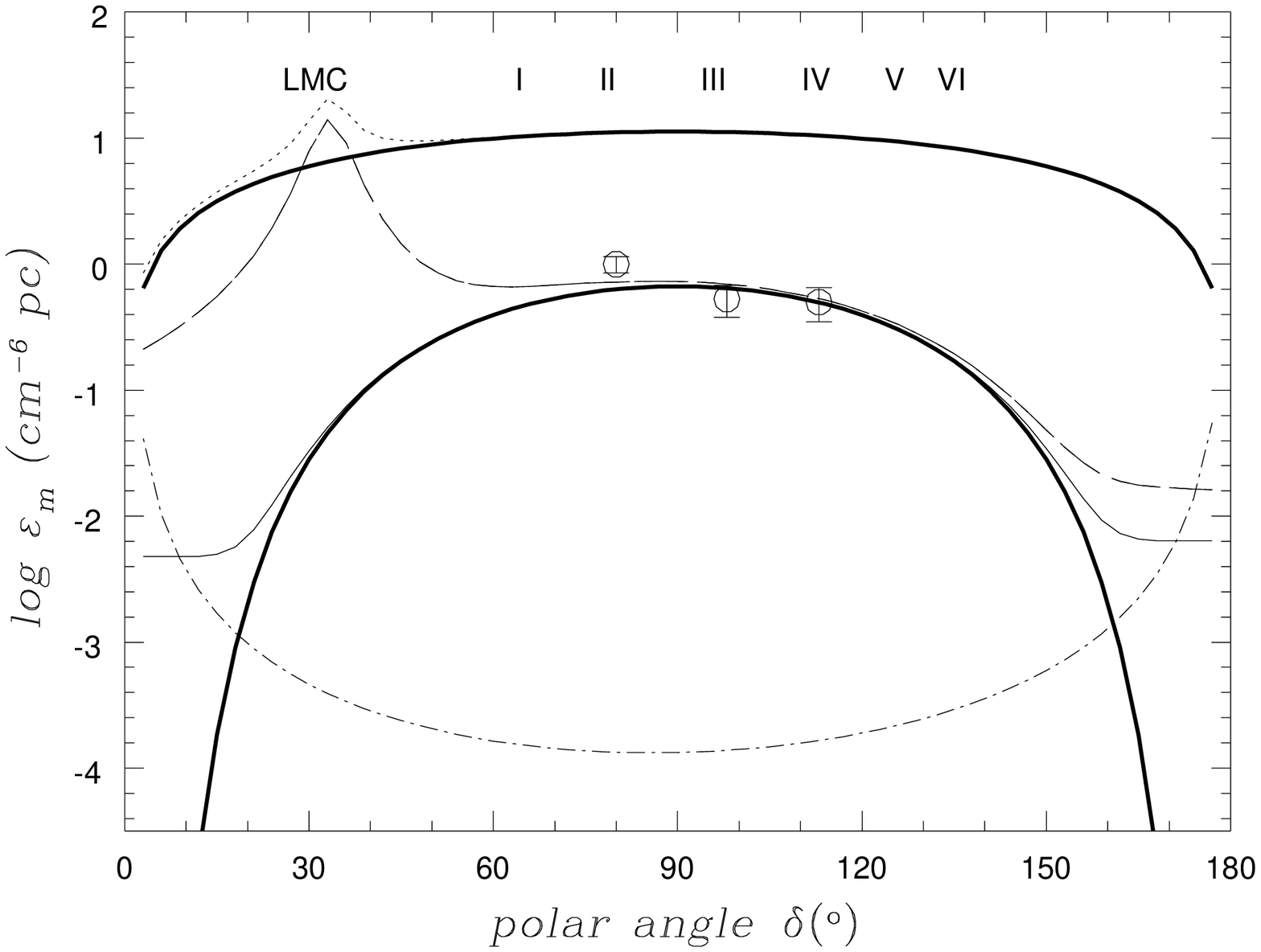,width=7cm}
\psfig{file=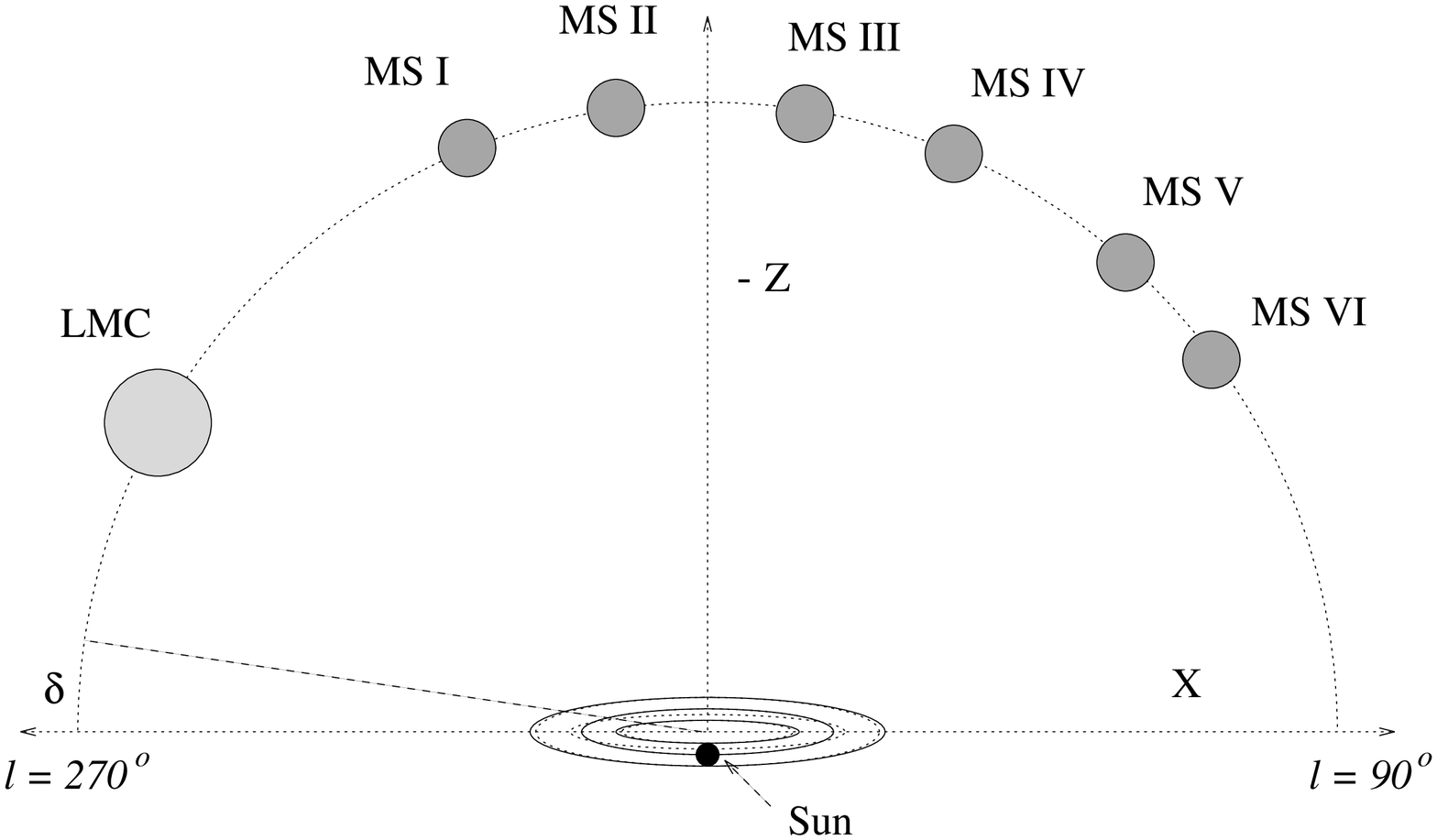,width=7cm}
}

\smallskip\noindent {\sl Fig.~5.} (a) The predicted emission measure 
along the Stream
as a function of $\delta$, defined in (b). The vertical axis has units of
log(cm$^{-6}$ pc), equivalent to log(Rayleighs) after subtracting
0.48.  The top curves assume an optically thin Galactic disk with (dashed
line) and without (solid line) the LMC ionizing field. For the lower
curves, the solid lines assume an opaque ($\tauLL = 2.8$) ionizing disk
with (thin line) and without (thick line) a bremsstrahlung halo; the LMC
contribution is shown by the long-dashed curve (using the same model as
Fig.~1). The dot-dash curve is $\Em$(\Ha) predicted for the upper side of the
Stream due to the bremsstrahlung halo; the cosmic ionizing background is
expected to dominate here. The open circles are the W$^2$ \Ha\ measurements. 

\medskip\bigskip
The basic ionizing requirement of the LMC from combined UV, optical and 
radio studies appears to be $5\times 10^{51}$\phoflux. Within a factor 
of two, this is consistent with OB star counts (Walborn 1984; Parker 1993), 
radio continuum observations (McGee, Brooks \& Batchelor 1972; 
Israel \& Koornneef 1979), and vacuum ultraviolet observations (Smith\etal\
1987) of the LMC. However, we note that the total number of ionizing photons 
produced by the LMC HII regions, spread over a 5 kpc region,
may be as high as $1.5-3\times 10^{52}\phorate$ (Smith\etal\ 1987). 
OB star counts around 30 Dor (\eg Parker 1993) could well underestimate the 
total ionizing flux by a substantial factor. 
Kennicutt\etal\ (1995) suggest that fully one third
of the ionizing radiation in the LMC arises from within 0.5\deg\ of 30 Dor.
The ground-based results may suffer 
from crowding which means that the total number of stars is underestimated. 

To examine this, we predict the \Ha\ emission along the Stream
using a high value for the intrinsic UV flux ($3\times 10^{52}\phorate$),
and assume that \fesc $=$ 50\%. In Fig.~5, \Ha\ observations of MS~I and MS~II 
are critical 
for determining the UV escape fraction from the LMC. The dilution of the 
radiation field should be detectable. This is crucial to determining the 3D 
orientation of the Leading Arm (Putman\etal\ 1998) in addition to the Stream.
A curious aspect of a substantial flux from the LMC is that this 
can mimic the appearance of shock ionization by lighting up the front 
surfaces of the Stream clouds.

If we assume that the trajectory of the
Magellanic Stream is described by a circular orbit with a radius of
roughly 50 kpc, then these clumps appear to lie close to the Galactic
polar axis extending to positive $X$ values.  Toomre (1972) first
suggested that the Clouds may have induced the observed warp in the
Galactic plane. But we note from Fig.~1 that the LMC could also have a
substantial ionizing effect on the cold gas in the plane of the HI
disk in the direction $l = 250^{\circ}-270^{\circ}$. In Fig.~1, the HI
disk warps towards negative $Z$ values along this axis (Burton 1988,
Fig. 7.23). It is plausible that the differences in the extent of the
warp along $l = 270^{\circ}$ compared with $l = 90^{\circ}$ are due to
ionization by the LMC, particularly if the escaping flux is at the
higher end of the quoted range.

\subsection{Spiral arms}

For HVC distances within 10 kpc of the plane, the halo ionization model needs 
a more realistic distribution of ionizing sources. This would be straightforward
if we knew the exact location of all O stars, and the precise dust distribution
throughout the Galaxy. At present, we assume that the disk UV field is
smoothly distributed; a more realistic model must include a 
predominant non-axisymmetric component.

Our updated `disk-halo' ionization model links its fortunes to the standard 
model for determining pulsar distances. Rough distances to pulsars are 
determined from the dispersion (and scattering) measure due
to warm electrons along the line of sight.  Early attempts used a smooth
distribution of electrons (\eg Manchester \& Taylor 1981) although 
Lyne, Manchester \& Taylor (1985) showed that typical distance estimates
have random errors as large as a factor of two.
After the inclusion of smooth spiral arms, Taylor \& Cordes (1993) predicted 
that most distances should be good to $\sim$20\%.  This level of accuracy
would be somewhat surprising when one examines face-on spirals in the 
Ultraviolet Imaging Telescope (UIT) database. But the distance model is largely
borne out by lower limits derived from pulsar sight lines which show HI in 
absorption.

The precise positions of spiral arms in the Galaxy are very difficult to 
determine. Early attempts to locate arms from continuous distributions (\eg HI)
are unreliable. The most reliable methods are those involving compact
sources identified from Galactic plane surveys, where follow-up observations
determine a velocity. A distance is inferred from the Galactic rotation
curve for sources that do not lie towards the Galactic centre. That this works 
fairly well is seen from comparing HII regions identified in Parkes single dish
measurements (with HI absorption velocities) with optical Fabry-Perot 
velocities (Georgelin \& Georgelin 1976).  Caswell \& Haynes (1987) find 
that the radio HII regions fall along the same loops in the (galactic 
longitude, LSR velocity) plane as the optical HII regions. The same spiral 
arms can be seen in the longitude$-$velocity diagram of giant molecular clouds
(Dame\etal\ 1987; Dame 1993; Digel\etal\ 1996).

Knowing the projected distribution of warm electrons is potentially useful.
Taylor \& Cordes (1993) note that 
\begin{quote}
{\sl Because the tangent directions to spiral arms are directly observable 
and independent of any distance estimates, they are 
defined without ambiguity. Maxima in the intensities of neutral hydrogen
and thermal radio emission occur at these longitudes precisely because large
quantities of interstellar gas, both neutral and ionized, are found along these
lines of sight.}
\end{quote}
While these can be different due to phase effects across spiral arms 
(\eg Rand 1998), the close association seen by Taylor \& Cordes suggests
that it may be possible to improve HVC distances close to the plane.
For our new model, we bypass the need to know either the O star or the dust
distribution. We assume only that the non-axisymmetric distribution of free 
electrons is proportional to the escaping UV flux from local sources.

The observed electron distribution from WHAM maps and pulsar dispersion 
measures reveal
a smooth component (assumed to be axisymmetric) and a non-axisymmetric 
component.  We normalize the total UV budget so that 6\% escapes into the halo
(Bland-Hawthorn \& Maloney 1999), 15$-$20\% escapes into the Reynolds layer
(Reynolds 1984; 1987; 1990; 1994), and some fraction of the total  
(20$-$60\%; Dahlem 1997; Dettmar 1992; Hoopes\etal\ 1996) is absorbed on 
scales of several hundred parsecs around the sites of star formation. 
(The quoted percentages of the total UV budget are not mutually exclusive.)
The remaining radiation
is reprocessed into low energy (non-ionizing) photons. Thus, we build up a 
more detailed disk-halo ionization model by scaling to the free 
electron density distribution of Taylor \& Cordes (1993).  
UIT observations indicate that the outer parts of mid to late type spirals 
look similar, although the inner parts exhibit a range of behaviour.

\section{Future issues}

There is a great deal to learn from future emission line (optical/IR)
studies of HVCs and HI structures in general. This conference has 
recognized the importance of obtaining \Ha\ measurements towards HI
clouds in order to derive accurate abundances, masses, etc.  
The authors have shown in a series
of papers that HI clouds should be observable well beyond 100 kpc from
the Galaxy. The radiation field is sufficiently strong to ionize HI
columns as high as 10$^{19}$ cm$^{-2}$. While as much as one third of
the sky is covered by HI, it would not be surprising if a similar 
fraction is in the form of H$^+$. 

A particularly exciting prospect is
to identify dense HI structures that have {\it no} \Ha\ counterpart
(\cf Blitz\etal\ 1999). Briggs (1998) and Staveley-Smith (1998, 
personal communication) note that stellar counterparts
are {\it always} seen in the vicinity of HI structures, even for
the famous Haynes-Giovanelli extragalactic cloud. Huge HI envelopes 
and tidal structures are observed in close association to galaxies
and groups of galaxies. It is surprising that more HI debris is
not found near clusters or galaxy groups, unless the gas becomes
dispersed and/or ionized. 

Finally, we have discussed the potential of \Ha\ detections for
obtaining distances to HVCs, and the pitfalls therein.  In the 
appendices that follow, we consider the prospect of seeing scattered
light from these same clouds. Indeed, at the risk of being incautious,
it is plausible that some HVCs could be pinpointed in 6-dimensional phase 
space (\S A.2). Elsewhere, we show that our AAT/WHT observing campaigns 
in early 1999 should uniquely distinguish between the competing dynamical 
models for the Stream (Bland-Hawthorn\etal\ 1999). A resolution 
of this longstanding controversy bears on many astrophysical problems.

\acknowledgments
JBH thanks toddler Christian Hawthorn for the squiggles in Fig.~3.
At the risk of ageism, we are indebted to J. Caswell for imparting 
several decades of experience in identifying HII regions at Parkes.  
R.N. Manchester, J.M. Chapman, M. Wardle and
S. Johnstone provided us with crucial insights.

\newpage
\bigskip
\noindent{\bf A.1: The Magellanic Stream seen in reflection?}

\medskip\noindent
An interesting possibility is that the W$^2$ \Ha\ detections arise
from the net \Ha\ emission of the Galaxy being backscattered by
high latitude dust. Such an explanation would appear to be unlikely.
The $R-$band to \Ha\ luminosity for an L$_*$ Sbc galaxy is in the
range 20 to 100, the clouds would be detectable in optical continuum,
which they are not to deep broadband limits (de Vaucouleurs 1954;
Recillas-Cruz 1982; Br\"{u}ck \& Hawkins 1983; Westerlund 1990). 
The estimated surface brightness is fainter than 27 mag arcsec$^{-2}$. 
As a further test, 
a dust-scattered line profile should show the imprint of the galaxy
rotation, increasingly so with decreasing galactic latitude. The observed 
line profiles (Weiner \& Williams 1996) appear to be relatively unresolved
although the MS clouds are at small polar angles (\ie large galactic
latitudes).  Standard dust models of the ISM predict that the grains are 
more forward scattering at bluer wavelengths
(Sokolowski, Bland-Hawthorn \& Cecil 1991, Fig. 3),
so it is unlikely that another spectral window will improve the chances
of detecting scattered light from the Magellanic Stream. However, high
velocity clouds closer to the Galactic plane are much more interesting, 
as we now discuss.

\bigskip\noindent
\psfig{file=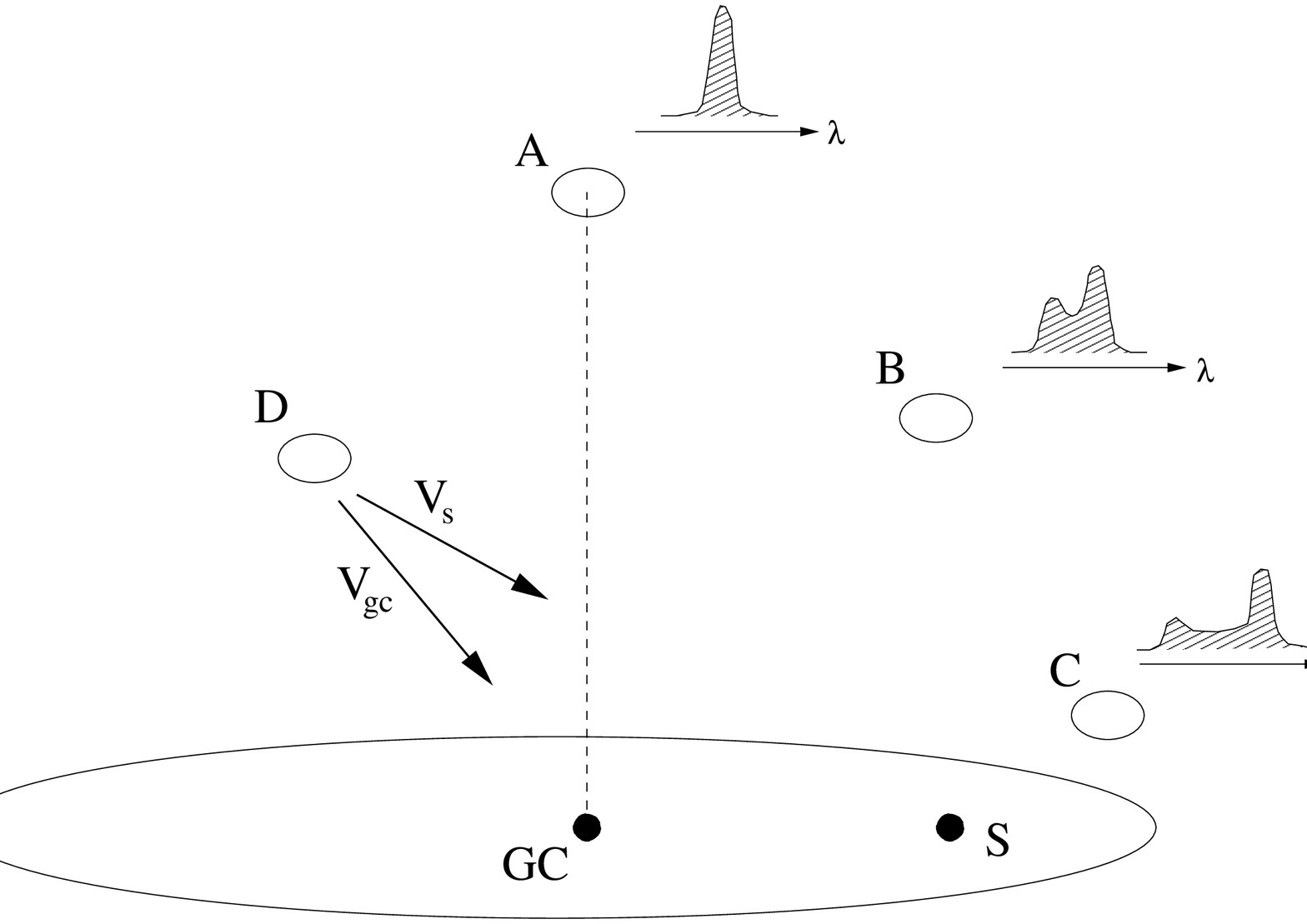,width=13cm}

\noindent {\sl Fig.~6.} Schematic twin horn \Ha\ profiles which could
arise from dust reflection in HI structures above the Galactic plane.
A simultaneous observation of \Ha\ due to ionization and \Ha\ seen
in reflection could be sufficient to pinpoint some HVCs in 6-dimensional 
phase space above the Galactic plane.

\bigskip
\noindent {\bf A.2: HVCs seen in reflection?}

\medskip\noindent
D.F. Malin (1998, private communication) has made a case for dust 
nebulosity observed towards some, presumably nearby, high latitude
clouds. If so, the emission observed in photographic plates surely 
arises from Galactic continuum light. This raises an intriguing possibility
(cf. Jura 1979).  Deep \Ha\ detections towards reflecting HVCs
could show emission from both ionization and reflection. The reflected
component will show twin horn structure (analogous to single beam HI
observations of spiral galaxies) from integrating over the Galactic 
disk, where the horn asymmetry and FWHM are directly related to the 
cloud's 3D position above the disk (see Fig.~6). The systemic velocity 
$\vec{v_s}$
of the twin horn structure will reflect both the motion of the dust 
mirror to the Galactic centre $\vec{v_{gc}}$ and the Sun $\vec{v_s}$, 
\ie $\vec{v_{o}} = \vec{v_{gc}} + \vec{v_s}$. If the \Ha\ flux due to
ionization is well separated from the reflected component, this could
conceivably allow us to locate some HVCs in 6-dimensional phase space
above the Galactic plane. Observed properties include the direction of 
a cloud in galactic coordinates, the horn FWHM, the degree of horn 
asymmetry, the horn offset from the HI velocity, and the \Ha\ ionization
flux.  Additional measurables include the twin-horn line flux and degree 
of linear polarization which would be difficult, but certainly not 
impossible, to measure. With these, the distance problem is now
overdetermined.

\end{document}